# Advances in Structure Prediction of Inorganic Compounds

Armel Le Bail - *Université du Maine, Laboratoire des oxydes et Fluorures, CNRS UMR 6010, Avenue O. Messiaen, 72085 Le Mans Cedex 9, France. Email : alb@cristal.org*

**Abstract** - There is a huge and confusing literature about inorganic crystal structure prediction. The word "prediction" is used sometimes as meaning "structure determination" since the process described needs the knowledge of the chemical composition and of the cell parameters. Some clarifications are presented here together with a new software (GRINSP) and some of its predictions.

## Introduction

To predict a crystal structure is not less than to be able to announce it before any confirmation by chemical synthesis or discovery in Nature. This seems to have little to do with powder diffraction. The relation becomes obvious only if a predicted structure is sufficiently accurate for the calculation of a predicted powder pattern that would further be used with success in the identification of a real compound not yet characterized. In a lead article entitled "Stuctural aspects of oxide and oxysalt crystals", Frank C. Hawthorne [1] stated, ten years ago: "The goals of theoretical crystallography may be summarized as follow: (1) predict the stoichiometry of the stable compounds; (2) predict the bond topology (i.e. the approximate atomic arrangement) of the stable compounds; (3) given the bond topology, calculate accurate bond lengths and angles (i.e. accurate atomic coordinates and cell dimensions); (4) given accurate atomic coordinates, calculate accurate static and dynamic properties of a crystal. For oxides and oxysalts, we are now quite successful at (3) and (4), but fail miserably at (1) and (2)". Surprisingly, four years earlier, it was stated that "computational methods can now make detailed and accurate predictions of the structures of inorganic materials" [2]. So, who shall we believe concerning inorganic predictions? The fact is that there are not a lot of predictions of inorganic compounds mentioned in the book edited by C.R.A. Catlow in 1997 [3] (note: concerning organic molecules, predictions do not appear more brillant, from the results of a recent blind test [4]).

If the state of the art had dramatically evolved in the past ten years, we should have huge databases of predicted compounds, and not any new crystal structure would surprise us since it would corespond already to an entry in that database. Moreover, we would have obtained in advance the physical properties and we would have preferably synthesized those interesting compounds. Of course, this is absolutely not the case. However, two databases of hypothetical compounds were built in 2003. One is exclusively devoted to zeolites [5], the other includes zeolites as well as other predicted oxides (phosphates, borosilicates, etc) and fluorides [6]. Some recent advances in inorganic crystal structure prediction are discussed in this manuscript.

### Previous knowledge versus prediction

Can we assume something to be known despite calling the process a "structure prediction" ? Some papers claiming for structure prediction assume the cell parameters and chemical composition to be known. In fact, in such a case we should classify this approach as a "structure determination technique". Having cell parameters means that one disposes of single crystal or powder diffraction data. Other works assume that the chemical composition only is known (at least tackling the packing problem). But if a real compound exists with that composition, then we do not have to predict its structure, we have to determine it by using diffraction techniques (single crystal or powder). Other assume that the bond valence rules will apply or that some selected energy potentials will provide the best cost function for driving the atom moves to a place corresponding to the convenient minimum of energy. These considerations are coming from our current knowledge of existing compounds, and it can appear justified to extrapolate such characteristics to still unknown compounds. It is expected from a prediction process that it will provide the structures of compounds to be synthesized or to be discovered in Nature, for which no chemical composition is known in advance, and of course no cell parameters. So, according to that definition, a few cases presented in the past as being predictions ($Li_3RuO_4$, $LiCoF_4$, $NbF_4$, etc) are in fact structure determinations realized by using more complex approaches than was really necessary. Several known methods for structure determination would not have had any difficulty to solve these problems either from powder diffraction or from single crystal data.

**Prediction software**

Let us cite shortly a few of the computer programs and methods producing predictions in the inorganic world. CASTEP uses the density functional theory (DFT) for ab initio modeling, applying a pseudopotential plane-wave code [7]. The structures gathered in the database of hypothetical zeolites [5] are produced from a 64-processor computer cluster grinding away non-stop, generating graphs and annealing them, the selected frameworks being then re-optimized using the General Utility Lattice Program (GULP, written by Julian Gale [8]) using atomic potentials. GULP itself appears to be able to predict crystal structures (one can find in the manual example 24 providing the data for the prediction of $TiO_2$ polymorphs). Recently, a genetic algorithm was implemented in GULP in order to generate crystal framework structures from the knowledge of only the unit cell dimensions and constituent atoms (so, this is not prediction...), the structures of the better candidates produced are relaxed by minimizing the lattice energy, which is based on the Born model of a solid [9] (this reference corresponds to a review about crystal structure prediction]. A concept of 'energy landscape' of chemical systems is used by Schön and Jansen for structure prediction [10] with their program named G42. Another package, SPuDS, is dedicated especially to the prediction of perovskites [11]. The AASBU method (Automated Assembly of Secondary Building Units) is developed by Mellot-Draznieks et al. [12], using Cerius2 [13] and GULP in a sequence of simulated annealing plus minimization steps for the aggregation of large structural motifs. GRINSP [14] applies the knowledge of the common geometrical characteristics of a well defined group of crystal structures (N-connected 3D nets with N = 3, 4, 5, 6 and combinations of two N values), in a Monte Carlo algorithm, allowing to explore the possible models, those already known (providing some proof of efficiency - see the table below) and those to be disclosed in a certain range of cell parameters. In GRINSP, the quality of a model is established by a cost function depending on the weighted differences between calculated and ideal interatomic first neighbour distances M-X, X-X and M-M in compounds $M_aX_b$ or $M_aM'_bX_c$. These models may need further optimization by using bond valence rules or larrice energy minimization, however, in many cases the predicted cell parameters differ by less than 2% from the real ones.

**Table 1 - Comparison of a few GRINSP-predicted cell parameters with observed ones**

| | Predicted (Å) | | | | Observed or idealized (Å) | | |
|---|---|---|---|---|---|---|---|
| Dense $SiO_2$ | a | b | c | R | a | b | c |
| Quartz | 4.965 | 4.965 | 5.375 | 0.0009 | 4.912 | 4.912 | 5.404 |
| Tridymite | 5.073 | 5.073 | 8.400 | 0.0045 | 5.052 | 5.052 | 8.270 |
| Cristobalite | 5.024 | 5.024 | 6.796 | 0.0018 | 4.969 | 4.969 | 6.926 |
| Zeolites | | | | | | | |
| ABW | 9.872 | 5.229 | 8.733 | 0.0056 | 9.9 | 5.3 | 8.8 |
| AFI | 13.836 | 13.836 | 8.514 | 0.0055 | 13.8 | 13.8 | 8.6 |
| ANA | 13.555 | 13.555 | 13.555 | 0.0025 | 13.6 | 13.6 | 13.6 |
| AST | 13.611 | 13.611 | 13.611 | 0.0059 | 13.6 | 13.6 | 13.6 |
| EAB | 13.158 | 13.158 | 15.034 | 0.0037 | 13.2 | 13.2 | 15.0 |
| EDI | 6.919 | 6.919 | 6.407 | 0.0047 | 6.926 | 6.926 | 6.410 |
| GIS | 9.772 | 9.772 | 10.174 | 0.0027 | 9.8 | 9.8 | 10.2 |
| GME | 13.609 | 13.609 | 9.931 | 0.0031 | 13.7 | 13.7 | 9.9 |
| JBW | 5.209 | 7.983 | 7.543 | 0.0066 | 5.3 | 8.2 | 7.5 |
| LTA | 11.936 | 11.936 | 11.936 | 0.0035 | 11.9 | 11.9 | 11.9 |
| MEP | 13.692 | 13.692 | 13.692 | 0.0077 | 13.7 | 13.7 | 13.7 |
| MER | 13.972 | 13.972 | 10.077 | 0.0026 | 14.0 | 14.0 | 10.0 |
| MON | 7.126 | 7.126 | 17.859 | 0.0052 | 7.1 | 7.1 | 17.8 |
| NAT | 13.822 | 13.822 | 6.414 | 0.0049 | 13.9 | 13.9 | 6.4 |
| RHO | 14.926 | 14.926 | 14.926 | 0.0022 | 14.9 | 14.9 | 14.9 |
| Aluminum fluorides | | | | | | | |
| τ-$AlF_3$ | 10.216 | 10.216 | 7.241 | 0.0162 | 10.184 | 10.184 | 7.174 |
| $Na_4Ca_4Al_7F_{33}$ | 10.860 | 10.860 | 10.860 | 0.0333 | 10.781 | 10.781 | 10.781 |
| $AlF_3$-pyrochlore | 9.668 | 9.668 | 9.668 | 0.0047 | 9.749 | 9.749 | 9.749 |

A problem with these predictions is the long calculation time. Those programs using trial and error procedures would benefit of parallel or grid computing. For instance, installed on a single processor PC running at 2GHz, the GRINSP software needs one day to examine one set of chemical elements in one space group, for random search of composition and random cell parameters (< 16 Å), so that the full exploration would need 230 days !

**More details on GRINSP**

**Generation of structure candidates** - With GRINSP, the building of the starting $M_aM'_bX_c$ model corresponds to a yes/no selection (the cost function is drastic) : first the M/M' atoms are placed in a box whose dimensions are selected at random, and the model should exactly correspond to the geometrical specifications (exact coordinations, but some tolerance on distances). The fact that distances are given a large tolerance range allows many solutions to be captured which may not correspond to regular polyhedra. In other words, the random walker may stay far above the deep local minima of interest. In this first step, atoms do not move, their possible positions are tested and checked, then they are retained or not. The cell is progressively filled up to completely respect the geometrical restraints, if possible. The number of M/M' atoms placed is not predetermined.

**Local optimization** - In a second step the X atoms are added at the midpoints of the (M/M')-(M/M') first neighbours and it is verified by distance and cell improvements (a Monte Carlo approach as well) that regular $(M/M')X_n$ polyhedra can really be built, i.e. that there is a deep local minima existing close to this previously selected rough arrangement of (M/M') atoms. The cost function allowing to establish a minimum is based on the verification of the provided ideal distances M-M, M-X and X-X first neighbours. The total R factor is defined as :

$$R = \sqrt{[(R_1+R_2+R_3)/(R_{01}+R_{02}+R_{03})]},$$

where $R_n$ and $R_{0n}$ for n = 1, 2, 3 are defined as :

$$R_n = \Sigma [w_n(d_{0n}-d_n)]^2,$$
$$R_{0n} = \Sigma [w_nd_{0n}]^2,$$

where $d_{0n}$ are the ideal first interatomic distances M-X (n=1), X-X (n=2) and M-M (n=3), whereas $d_n$ are the observed distances in the structural model. The weights retained ($w_n$) are those used in the DLS [15] software for calculating idealized zeolite framework data ($w_1$= 2.0, $w_2$ = 0.61 and $w_3$ = 0.23). The ideal distances are to be provided by the user for pairs of atoms supposed to form polyhedra (for instance in the case of $SiO_4$ tetrahedra, one expects to have $d_1$ = 1.61 Å, $d_2$ = 2.629 Å and $d_3$ = 3.07 Å). The similarity of the cell parameters estimated by GRINSP for zeolites with the idealized cell constant listed at the official zeolite Web site [16] is not fortuitous, since these idealized values are calculated by using the DLS software applying a similar cost function during the distance least square refinements. For ternary compounds, the M-M' ideal distances are calculated by GRINSP as being the average of the M-M and M'-M' distances. It is clear that this R factor considers only the X-X intra-polyhedra distances, neglecting any X-X inter-polyhedra distances This cost function R could possibly be better defined differently, for instance by using the bond valence sum rules (this is in project for the next GRINSP version). During this second step, the atoms are moving, but no jump is allowed because a jump would break the coordinations established at the first step. This is a simple routine for local optimization. The change in the cell parameters from the structure candidate to the final model may be quite considerable (up to 30%), this explains why some models may show parameters larger or smaller than limits defined during the runs, these limits being applied only to the first step results  During the optimization, the original space group used for placing the M/M' atoms may change after adding the X atoms, so that the final structure is always proposed in the P1 space group, and presented in a CIF file. The final choice of the real symmetry has to be done by using a program like PLATON. One given model can be retrieved in different space groups with sligthly different R values. One can imagine using a parallel computer with a GRINSP version which would allow also to select randomly the space group, so that one global run would provide the optimal space group for each structure type, the best results being sorted out only at the end of the process.

**Binary compounds predicted by GRINSP**

Formulations $M_2X_3$, $MX_2$ and $MX_3$ were partly examined (not yet $M_2X_5$ which would occur for M cations in fivefold coordination).

**Zeolites -** The complete exploration is still not finished. A thousand models are expected to come from GRINSP with R < 0.01 and cell parameters < 16 Å. The PCOD database contains already more than 300 models mainly in cubic and hexagonal symmetry. Examples establishing the quality of the predictions are presented in Table 1 showing already known zeotypes retrieved by the program. The way GRINSP recognizes a zeotype is by comparing the coordination sequence (CS) [17] of any model with a list of previously established ones (as well as with the other CS already stored during the current run). A few of these hypothetical zeolites with small framework density (FD : number of Si atoms for 1000 Å$^3$) are presented below, one ordered Si/Al prediction being among them. The CIF files can be obtained by consulting the PCOD database [6], giving the entry number provided with the figure caption (for instance PCOD1010026, etc).

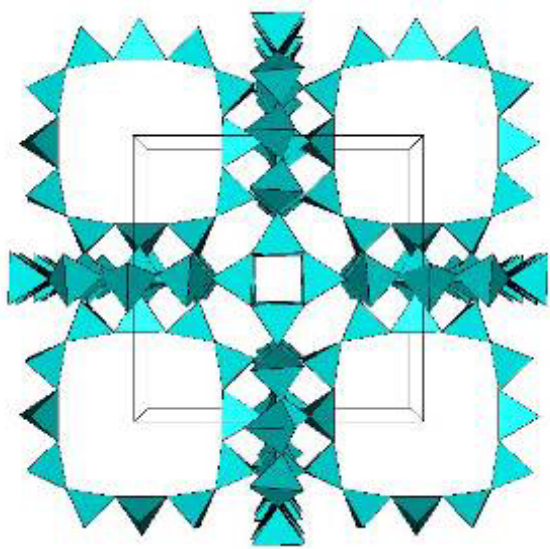

Hypothetical zeolite PCOD1010026
SG : P432, a = 14.623 Å, FD = 11.51

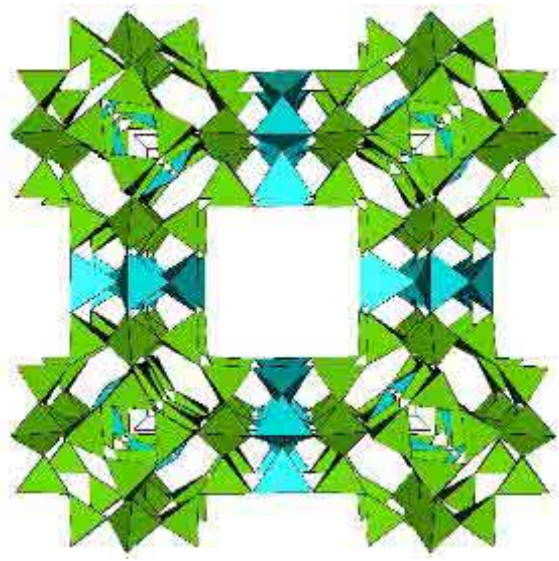

Hypothetical zeolite PCOD1010038
SG : P432, a = 14.70 Å - FD = 11.32
formulation : $[Si_2AlO_6]^{-1}$

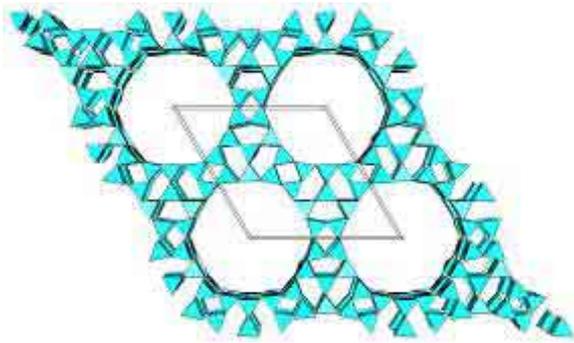

Hypothetical zeolite PCOD1030060
SG : P6$_3$mc, a = 18.74 Å, c = 9.02Å, FD = 14.6.

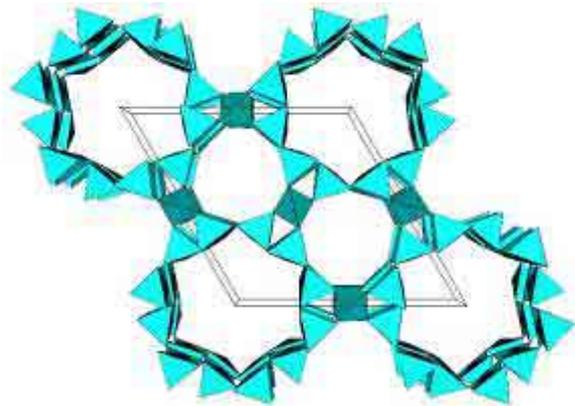

Hypothetical zeolite PCPD1030067
SG : P6cc, a = 14.48 Å, c = 9.17Å, FD = 18.0.

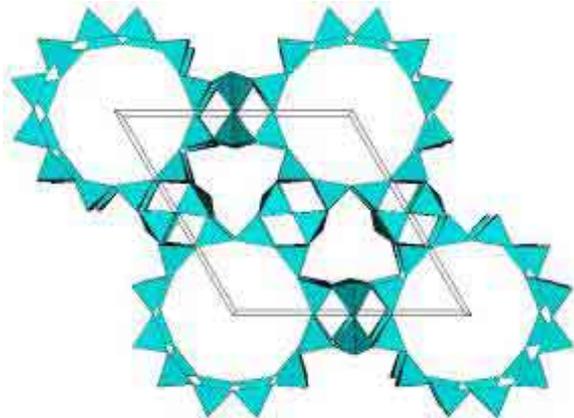

Hypothetical zeolite PCOD1030081
SG : P6/m, a = 15.60 Å, c = 7.13Å, FD = 16.0.

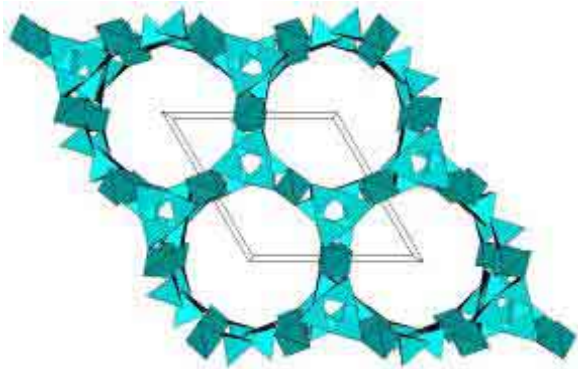

Hypothetical zeolite PCOD1030083
SG : P 6cc, a = 14.14 Å, c = 13.90 Å, FD = 17.4.

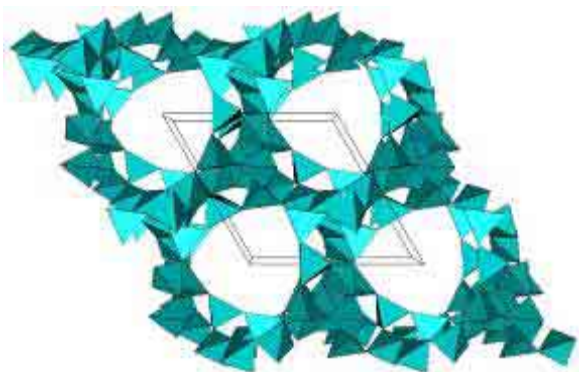

Hypothetical zeolite PCOD1030157
SG P3$_2$21, a = 11.18 Å, c = 11.24Å, FD = 17.3.

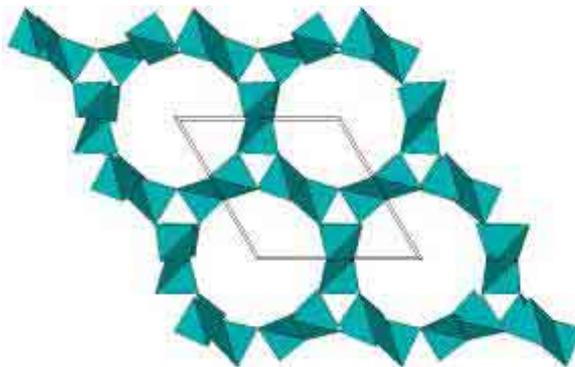

Hypothetical zeolite PCOD1030129
SG : P6１, a = 8.661 Å, c = 4.403Å, FD = 21.0.

**B₂O₃ polymorphs predicted by GRINSP -** Not a lot of crystalline varieties are known for this B₂O₃ composition. Too many are proposed by GRINSP. A complete exploration may not be justified due to some lack of interest ?

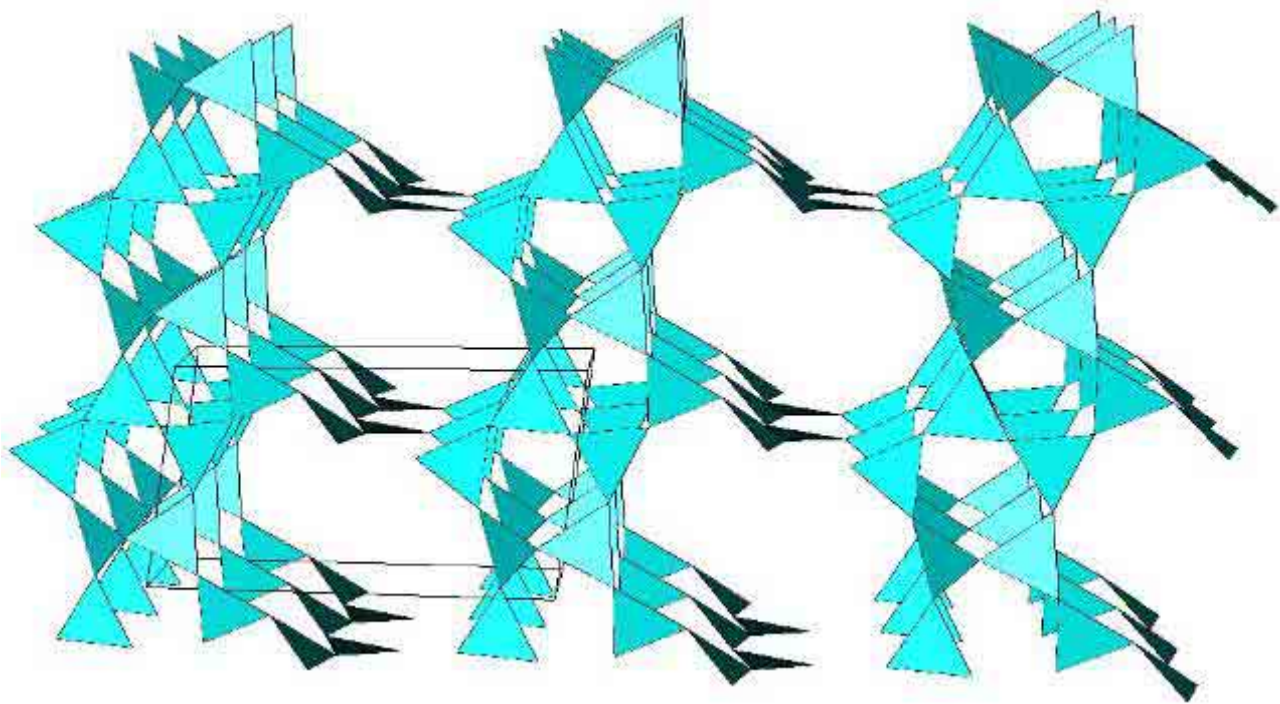

Hypothetical B₂O₃ PCOD1062004.

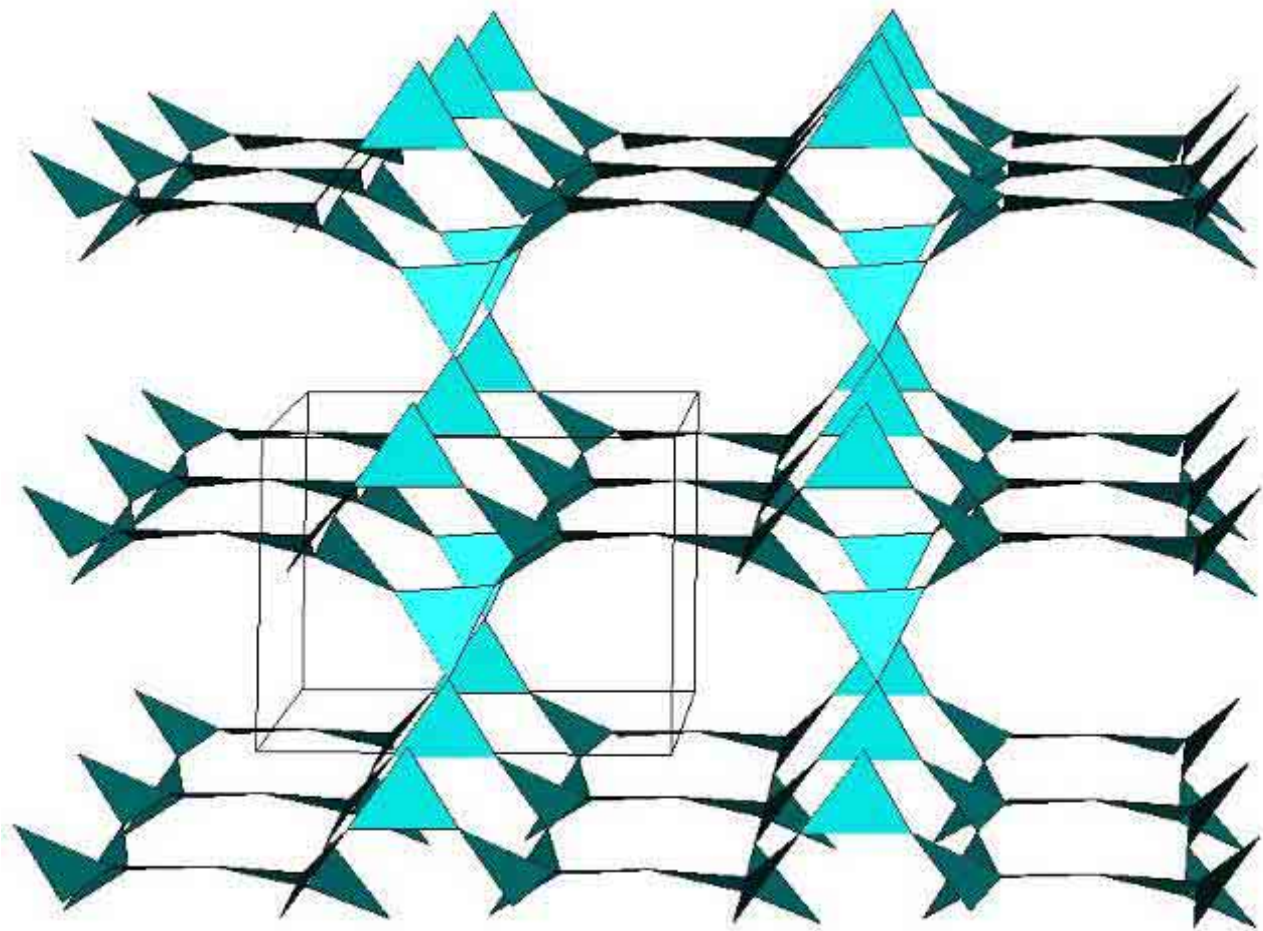

Hypothetical B₂O₃ PCOD1051002.

**AlF$_3$ polymorphs yet to be synthesized, predicted by GRINSP** - Apart from the well known perovskite structure type, which can be retrieved in almost all space groups during the exploration of the 6-connected 3D nets with GRINSP, all the known structure-types were retrieved, including that of τ-AlF$_3$ [18] A series of "yet to be synthesized" AlF$_3$ polymorphs were also proposed (which apparently were not disclosed by the AASBU process [12], the known τ-AlF$_3$ structure type itself being not mentioned). Some structure types are really known with AlF$_3$ formulation, or at least may exist with other MX$_3$ formulations, stuffed or not (K$_x$FeF$_3$, etc). The complete search was made in the 230 space groups, however it was restricted to cell parameters smaller than 16 Å, using the following first-neighbours ideal interatomic distances : 3.5 Å for Al-Al, 1.81 Å for Al-F and 2.559 Å for F-F in the R calculations. The range of distances selected at large for obtaining the initial models with Al only atoms were 2.90-4.10 Å for first Al-Al neighbours and 4.20-6.70 Å for the second Al-Al neighbours. The F atoms being then added at the midpoints of the Al-Al first neighbours, and the model being Monte Carlo refined up to obtain regular octahedra. Due to the Monte Carlo approach, models may have escaped the search which was limited to 10000-200000 tests per space group, allowing 300000 MC events (for positioning first the Al atoms) inside of each test. For the model improvement in the second step, up to 10000-20000 MC events (either moving Al of F atoms or changing the cell parameters) were allowed.

**Table 2 - Classification of the AlF$_3$ polymorphs proposed by GRINSP (identified as known or unknown) according to increasing values of the distance quality factor R**

| Structure-type | FD | a | b | c | α | β | γ | SG | Z | N | R |
|---|---|---|---|---|---|---|---|---|---|---|---|
| HTB | 19.68 | 6.987 | 6.987 | 7.212 | 90.00 | 90.00 | 120.00 | P6$_3$/mmc | 6 | 1 | 0.0035 |
| TlCa$_2$Ta$_5$O$_{15}$ | 20.67 | 7.004 | 7.228 | 9.558 | 90.00 | 90.00 | 90.00 | Pmmm | 10 | 2 | 0.0040 |
| U-1 (AlF$_3$) | 21.27 | 6.992 | 7.218 | 13.513 | 90.00 | 105.22 | 90.00 | P2$_1$/m | 14 | 3 | 0.0042 |
| Pyrochlore | 17.71 | 9.668 | 9.668 | 9.668 | 90.00 | 90.00 | 90.00 | Fd-3m | 16 | 1 | 0.0046 |
| U-2 (AlF$_3$) | 20.43 | 6.889 | 6.889 | 8.252 | 90.00 | 90.00 | 90.00 | P-4m2 | 8 | 2 | 0.0057 |
| Perovskite | 21.16 | 3.615 | 3.615 | 3.615 | 90.00 | 90.00 | 90.00 | Pm-3m | 1 | 1 | 0.0063 |
| Ba$_4$CoTa$_{10}$O$_{30}$ | 21.15 | 9.499 | 13.777 | 7.224 | 90.00 | 90.00 | 90.00 | Iba2 | 20 | 2 | 0.0095 |
| TTB | 20.78 | 11.539 | 11.539 | 7.229 | 90.00 | 90.00 | 90.00 | P4$_2$/mbc | 20 | 2 | 0.0099 |
| U-3 (AlF$_3$) | 22.37 | 6.960 | 7.402 | 5.207 | 90.00 | 90.00 | 90.00 | Pnc2 | 6 | 2 | 0.0160 |
| τ-AlF$_3$ | 21.17 | 10.214 | 10.214 | 7.242 | 90.00 | 90.00 | 90.00 | P4/nmm | 16 | 3 | 0.0162 |
| U-4 (AlF$_3$) | 21.71 | 10.505 | 10.505 | 6.678 | 90.00 | 90.00 | 90.00 | I4$_1$/a | 16 | 1 | 0.0181 |
| U-5 (AlF$_3$) | 19.74 | 7.125 | 7.125 | 11.977 | 90.00 | 90.00 | 90.00 | P4$_2$/mmc | 12 | 2 | 0.0191 |
| U-6 (AlF$_3$) | 23.65 | 12.601 | 12.601 | 6.391 | 90.00 | 90.00 | 90.00 | P4/nmm | 12 | 2 | 0.0233 |
| U-7 (AlF$_3$) | 19.22 | 6.396 | 6.396 | 5.087 | 90.00 | 90.00 | 90.00 | P4$_2$mc | 4 | 1 | 0.0243 |
| U-8 (AlF$_3$) | 19.65 | 10.624 | 10.624 | 7.212 | 90.00 | 90.00 | 90.00 | P4/mmm | 16 | 2 | 0.0275 |
| Na$_4$Ca$_4$Al$_7$F$_{33}$ | 23.27 | 9.805 | 9.805 | 9.834 | 90.00 | 90.00 | 90.00 | I4/mmm | 22 | 3 | 0.0283 |
| U-9 (AlF$_3$) | 23.46 | 7.997 | 7.997 | 7.997 | 90.00 | 90.00 | 90.00 | P4$_1$32 | 12 | 2 | 0.0287 |
| U-10 (AlF$_3$) | 17.68 | 6.874 | 6.874 | 14.360 | 90.00 | 90.00 | 90.00 | P42mc | 12 | 2 | 0.0299 |

FD = framework density (number of Al atoms for a volume of 1000Å$^3$).
SG = higher symmetry spage group in which the initial model of Al-only atoms was obtained (not being necessarily the true final space group obtained after including the F atoms).
Z = number of AlF$_3$ formula per cell.
N = number of Al atoms with different coordination sequences.
R = quality factor regarding the ideal Al-F, F-F and Al-Al first neighbour interatomic distances.

Up to R < 0.02, five unknown structure types are disclosed by GRINSP which could well constitute some viable "yet to be synthesized" AlF$_3$ compounds. Three of them (noted U-1, U-2 and U-3) have even R values smaller than for the known metastable compound τ-AlF$_3$ (R = 0.0162), and the two others (noted U-4 and U-5) present R values only slightly higher than 0.0162. U-1 is a simple HTB-perovskite intergrowth with one more perovskite layer than into the TlCa$_2$Ta$_5$O$_{15}$ intergrowth structure type. Tetrahedra of octahedra, like in the pyrochlore or the τ-AlF$_3$ structure types are recognized also in the U-2 and U-4 models. For R values larger than 0.02, problems may arise like too short interatomic distances for a small part of them, and some octahedra are becoming much more distorted. For instance, in spite of an interesting channel with rings of 8 octahedra, U-6 has a large framework density because of the too close proximity of the octahedra along the c axis (distances Al-Al = 3.195 Å, really too short for corner sharing). Other models listed with 0.02 < R < 0.03 are probably not viable with this AlF$_3$ formulation but could be encountered as MX$_3$ compounds if the MX$_6$ octahedra accept some distorsion. Only one (U-10) of these models presents a framework density slightly smaller than the pyrochlore one, but some too short Al-F distances are certainly prohibiting its existence. The model showing the Na$_4$Ca$_4$Al$_7$F$_{33}$ arrangement replaces the Ca atoms by Al ones, this model is also obtained among the ternary compounds with a [Ca$_4$Al$_7$F$_{33}$]$^{4-}$ 6-connected 3D network, but GRINSP is still unable to place the Na atoms in the holes. The figures below are corresponding to 12 of the 13 cases with R values up to 0.0233 (the simple well known perovskite being excluded).

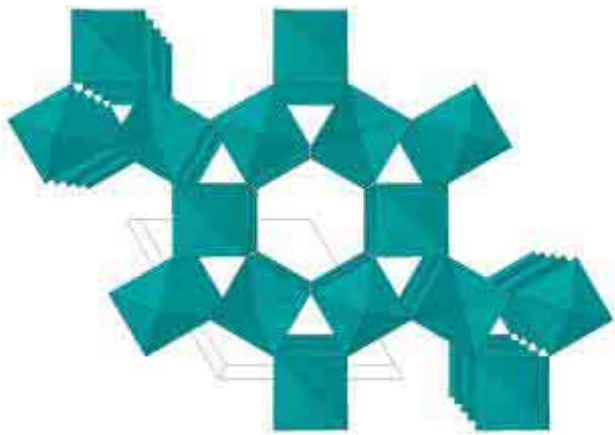

HTB (Hexagonal Tungsten Bronze) structure-type

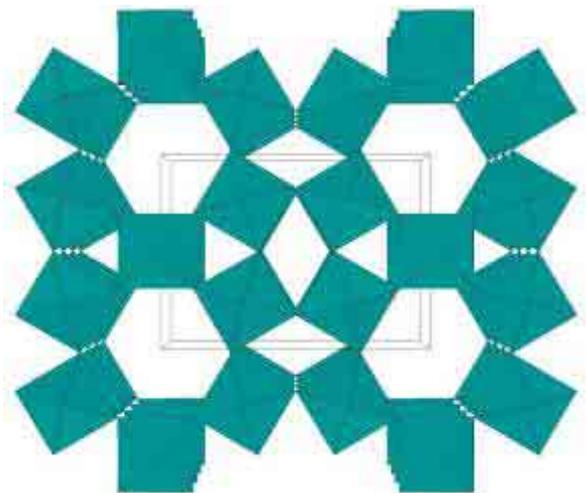

TlCa$_2$Ta$_5$O$_{15}$ structure-type,
intergrowth HTB-perovskite (2 layers)

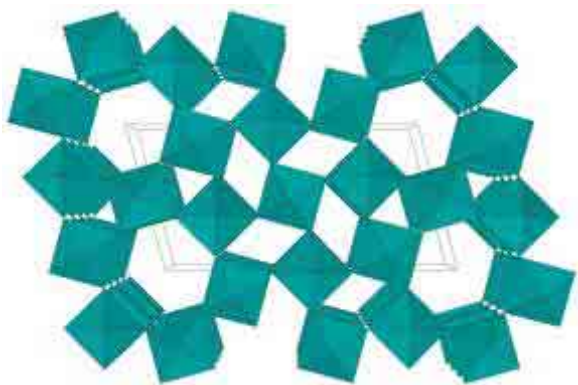

Yet to be synthesized U-1,
intergrowth HTB-perovskite (3 layers)

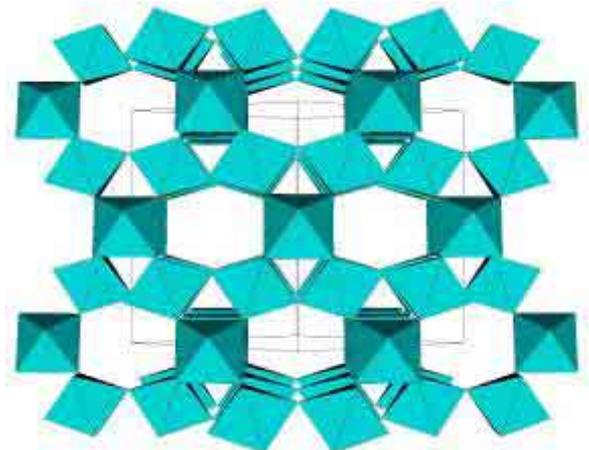

Pyrochlore structure-type,
built up from tetrahedra of octahedra

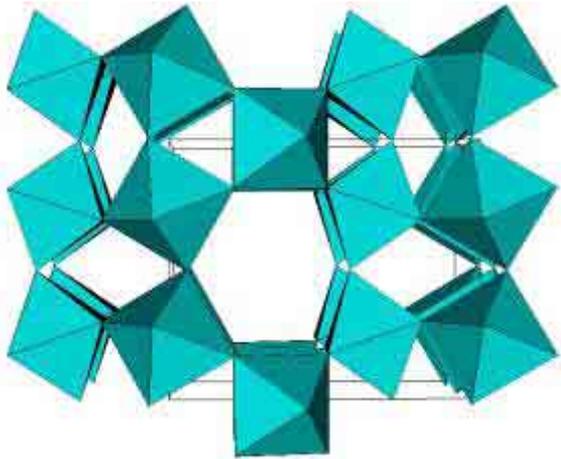

Yet to be synthesized U-2 (AlF$_3$), intergrowth pyrochlore-perovskite

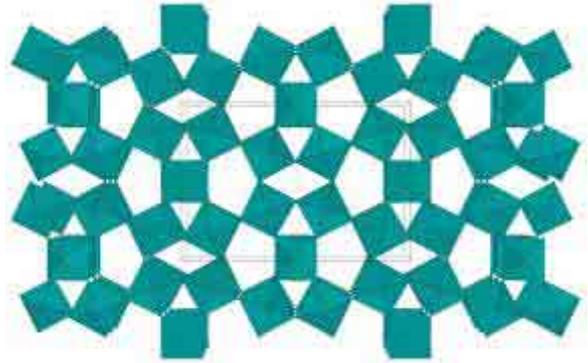

Ba$_4$CoTa$_{10}$O$_{30}$ structure-type

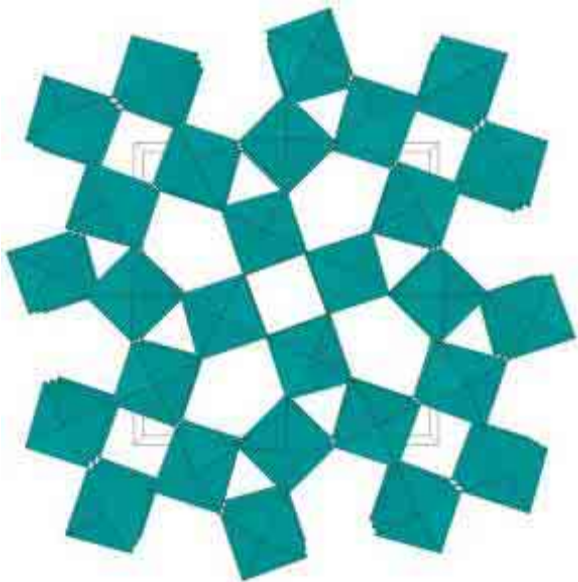

TTB structure-type
(Tetragonal Tungsten Bronze)

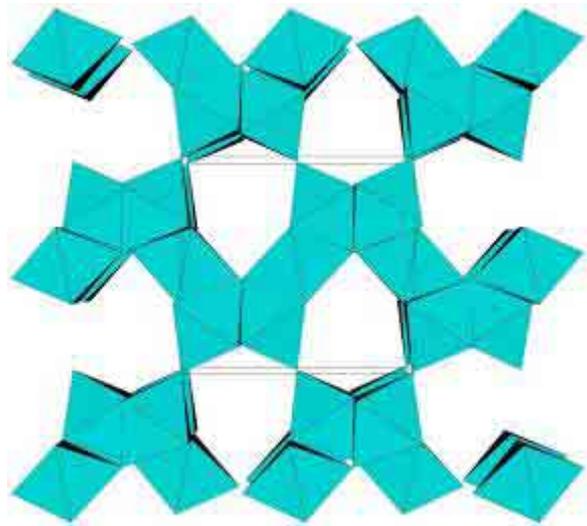

Yet to be synthesized U-3 (AlF$_3$).

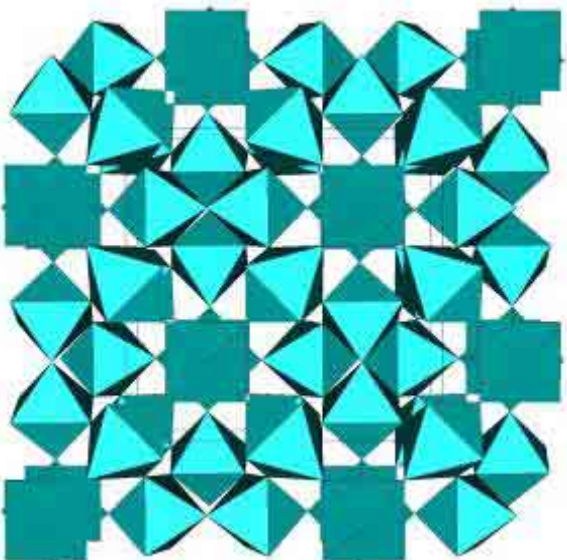

τ-AlF$_3$ - tetrahedra and chains of octahedra

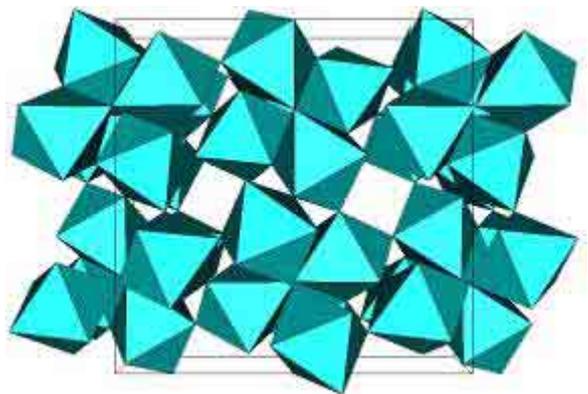

U-4 (AlF$_3$), dense packing of tetrahedra of octahedra, exclusively

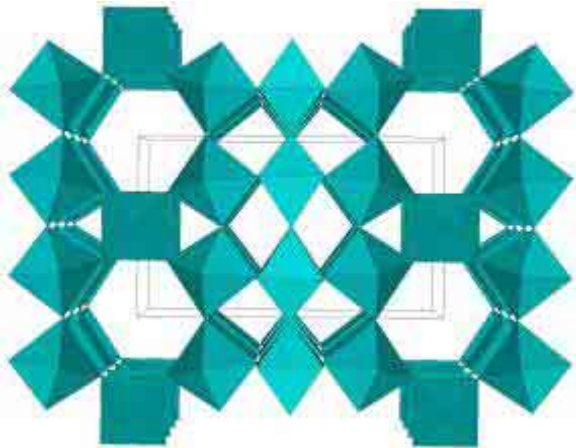

U-5 (AlF$_3$), HTB tunnels intercrossed at 90° in the ab plane

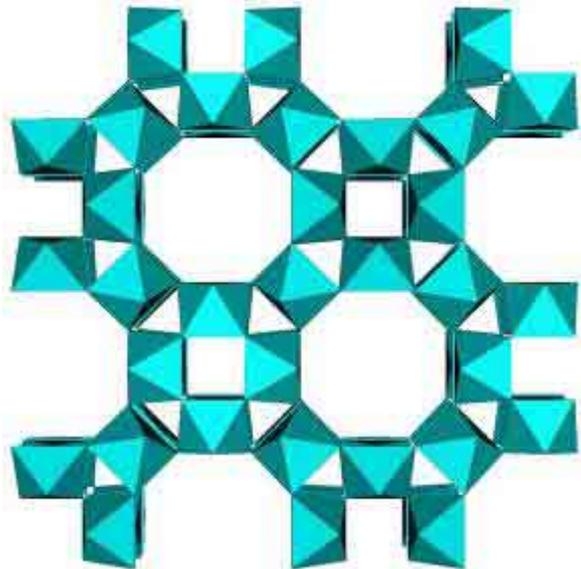

U-6 (AlF$_3$), not viable due to a too high level of octahedra distortion and short F-F distances

**By-products of the search with GRINSP -** Other sixfold polyhedra than octahedra can be produced: trigonal prisms or pentagonal based pyramids. Since they do not correspond to one unique ideal F-F distance or Al-F distance, they are ranked with high R-values. Aluminum is not known in fluoride solids with other coordination than a very regular octahedron, so that such predictions are very probably useless. However, on the point of view of the structures, some were surprisingly presenting very small framework densities and may be of interest. Two examples are shown below.

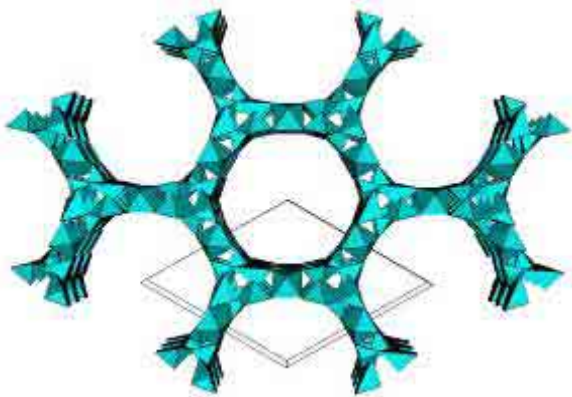
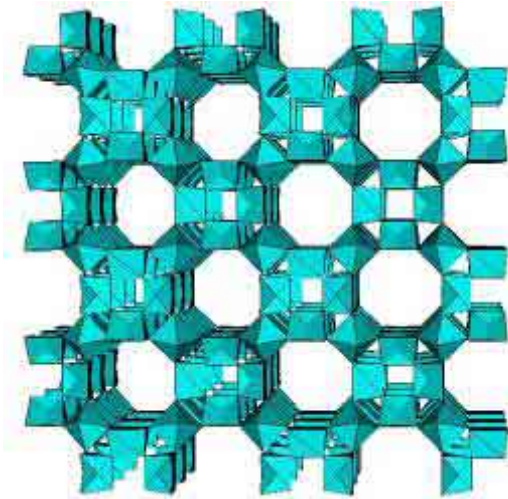

Moreover, many two-dimensionnal compounds can be formed which will correspond to all polyhedra corners satisfied. In such cases, GRINSP has no way to make any correct estimate of the intersheet distance, so that these models were not collected (they will possibly correspond to extremely small FD values). Some one-dimensionnal models were even built (cylinders with B$_2$O$_3$ formulation for instance). Also in that case, the distances between the rods could not be estimated and the cell parameters are fanciful. A picture of these B$_2$O$_3$ cylinders is represented below.

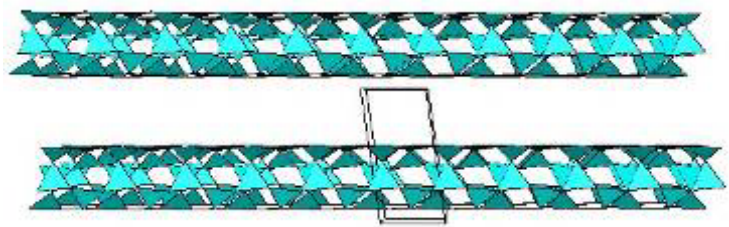

**Ternary compounds with corner-sharing 3D nets**

Here, M and M' cations are considered. They could have a same coordination but different ionic radii (allowing to explore ordered aluminosilicates or phosphosilicates) or different coordination, but the current limitation with GRINSP is that the connections by X atoms will only be by corner sharing: all X atom should be connected to exclusively two M atoms or two M' atoms or one M and one M' atom. As a consequence, only some formulations can occur which fulfill these conditions, moreover, if M or M' are not able to form electrically neutral binary compound with corner-sharing only, then the built ternary compound will also not be electrically neutral. All the borosilicates formed with GRINSP are automatically electrically neutral since B is involved in $BO_3$ triangles corner sharing $B_2O_3$ polymorphs and Si occurs in $SiO_4$ tetrahedra corner sharing $SiO_2$ dense polymorphs and zeolites

**Borosilicates** - There is only one hit in the ICSD database for this kind of compound. A strange result is that GRINSP produces a huge quantity of hypothetical borosilicates, showing exclusively $BO_3$ triangles and $SiO_4$ tetrahedra linked by corners. Limiting $R < 0.006$, working in cell symmetry higher or equal to monoclinic, but using the general Wyckoff position of the P1 space group, 57 different models were found with $SiB_2O_5$ formulation, 32 models for $Si_3B_4O_{12}$, 28 for $Si_2B_6O_{13}$ and $Si_4B_2O_{11}$, 24 for $Si_2B_2O_7$, 18 for $SiB_6O_{11}$, 17 for $SiB_4O_8$, 14 for $Si_3B_2O_9$, 6 for $Si_6B_2O_{15}$ and 2 for $Si_3B_6O_{15}$. Moreover, 369 different additional models were disclosed in triclinic symmetry ! The number of these models would probably explode by a complete search in the 230 space groups since the introduction of Wyckoff positions having more than one equivalent boosts the capacity of the GRINSP software having difficulties to find structures more complex than 10-20 independent M/M' atoms in a triclinic cell. Those hypothetical borosilicates are not all yet enclosed into the PCOD. Two of them are shown below.

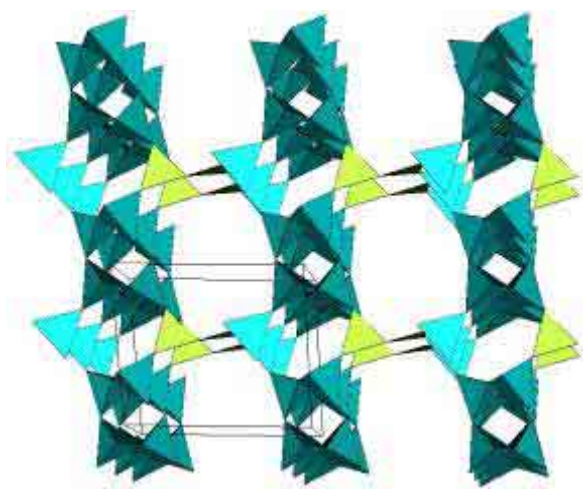 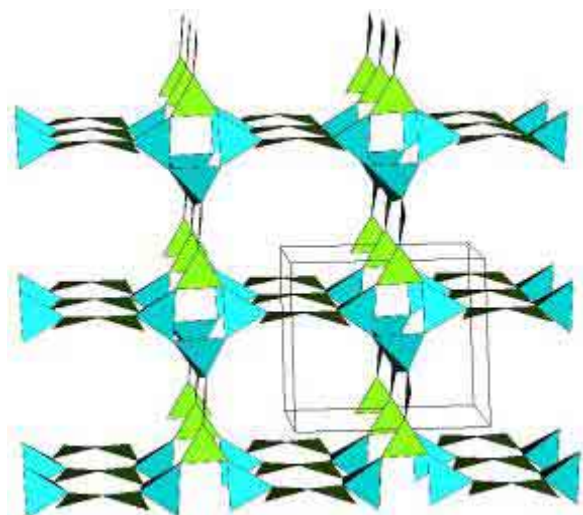

PCOD2050102, $Si_5B_2O_{13}$, R = 0.0055.    PCOD2050018, $Si_3B_4O_{12}$, R = 0.0039.

**Titanosilicates** - Explorations in this domain (in fact a part of the domain where octahedra and tetrahedra are exclusively corner-linked) are in progress. The number of hypothetical strutures with small R values is large and the enumeration is far from being finished. Many structure-types existing for other compositions were enumerated. A very small part of the new models is proposed below, some being clearly not viable if the polyhedra do not accept some distorsion (those models have R > 0.02). The models are not electrically neutral so that the frameworks would have to accept some additional cations or charged molecule for existing really. A next step in the GRINSP development is clearly to add the automatic filling of holes by cations able to bring neutrality.

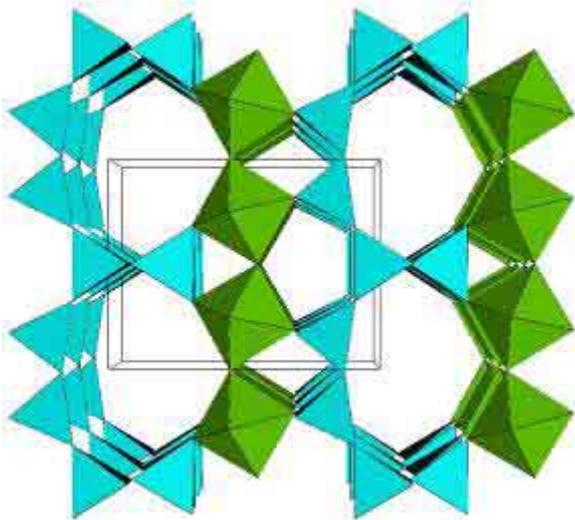

Hypothetical titanosilicate $[Si_6Ti_4O_{22}]^{8-}$, R = 0.0101
SG : P-4m2, a = 7.564 Å, c = 9.702 Å, FD = 16.2.

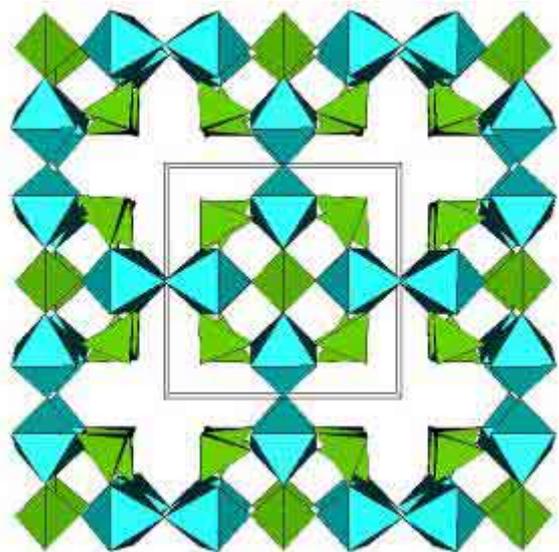

Hypothetical titanosilicate $[Si_9Ti_4O_{30}]^{4-}$, R = 0.0181
SG : P4/mmm, a = 10.19 Å, c = 5.46 Å, FD = 22.9.

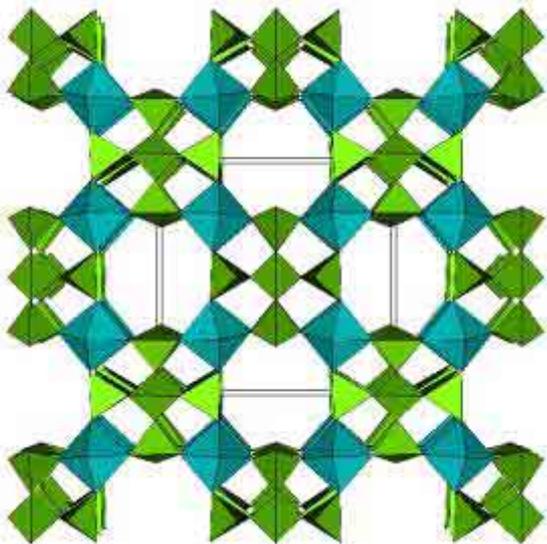

Hypothetical titanosilicate $[Si_5Ti_2O_{16}]^{4-}$, R = 0.0132
SG : I4/mmm, a = 13.12 Å, c = 7.69 Å, FD = 21.1.

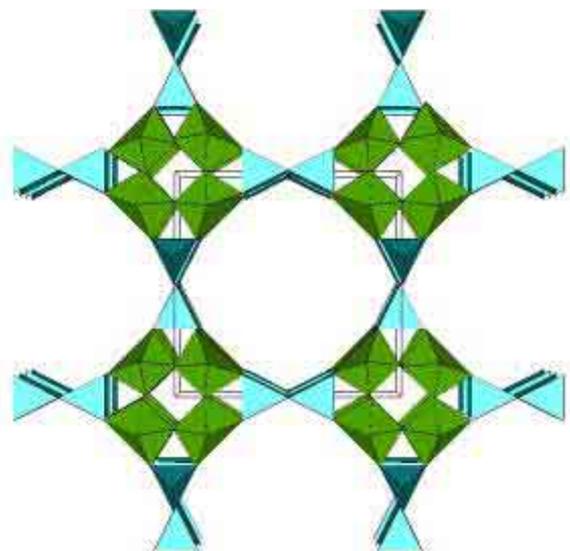

Hypothetical titanosilicate $[SiTiO_5]^{2-}$, R = 0.0335
SG : P4$_2$/mmc, a = 12.85 Å, c = 7.76 Å, FD = 12.5.

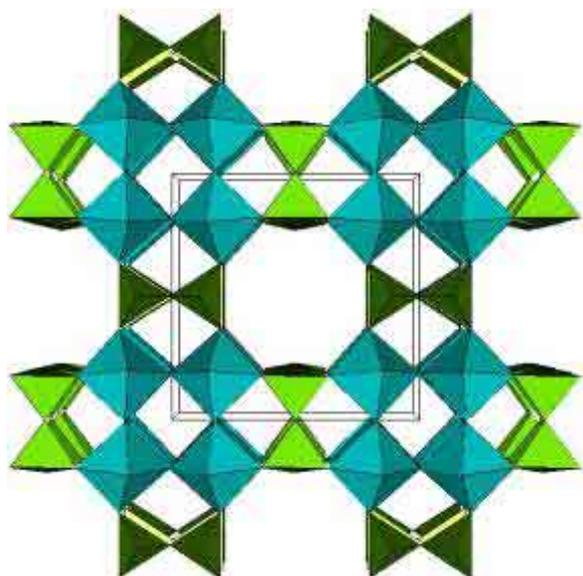

Hypothetical titanosilicate $[SiTiO_5]^{2-}$, R = 0.0109
SG : P4$_2$/mmc, a = 10.47 Å, c = 7.64 Å, FD = 19.1.

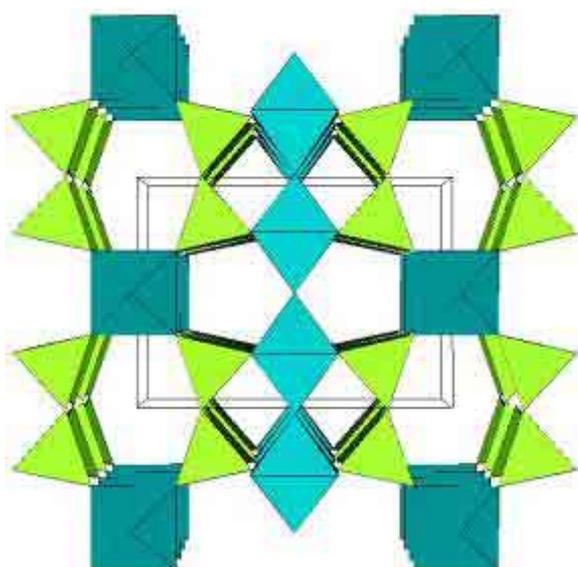

Hypothetical titanosilicate $[Si_2TiO_7]^{2-}$, R = 0.0044
SG : P4$_2$/mmc, a = 7.73 Å, c = 10.50 Å, FD = 19.1.

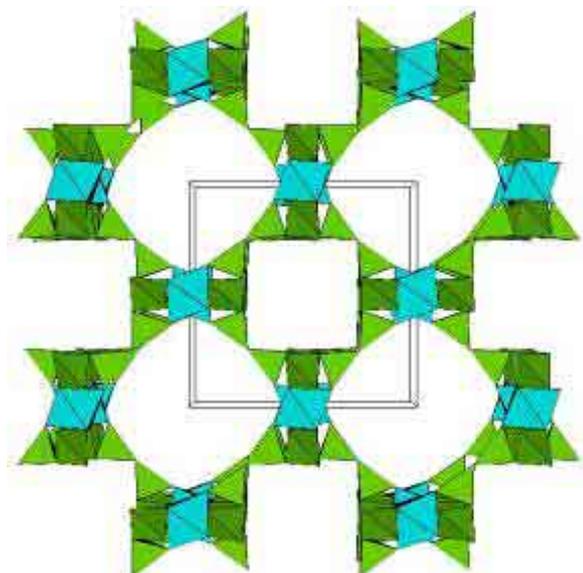

Hypothetical titanosilicate $[Si_4TiO_{11}]^{2-}$, R = 0.0175
SG : P4$_2$/mmc, a = 12.89 Å, c = 9.23 Å, FD = 13.0.

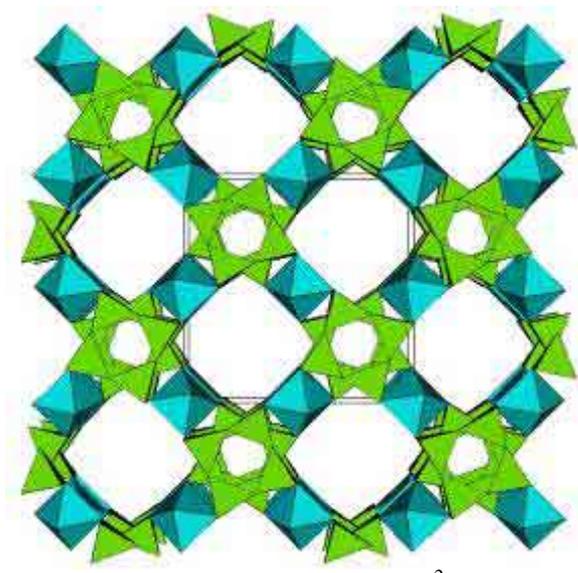

Hypothetical titanosilicate $[Si_2TiO_7]^{2-}$, R = 0.0165
SG : P4$_2$/nbc, a = 12.32 Å, c = 7.27 Å, FD = 21.7.

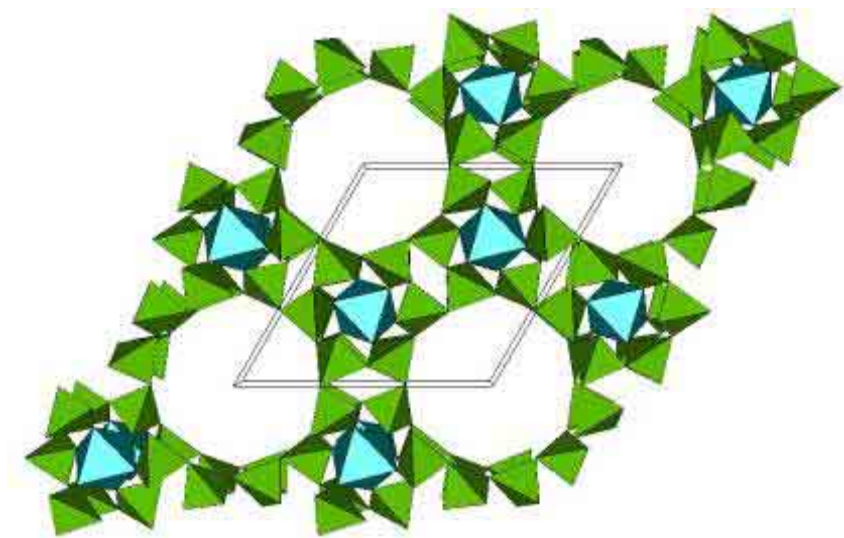

Hypothetical titanosilicate $[Si_6TiO_{15}]^{2-}$, R = 0.0124,
SG : P6$_3$, a = 11.97 Å,
c = 6.30 Å, FD = 17.9.

**Fluoroaluminates** - This category of compounds was only partly explored up to now, combining octahedra with different sizes (AlF$_6$ and CaF$_6$ or NaF$_6$). Some known 6-connected frameworks were retrieved, such as [Ca$_4$Al$_7$F$_{33}$]$^{4-}$ existing as Na$_4$Ca$_4$Al$_7$F$_{33}$. A model was obtained too replacing Ca by Al but the R factor was high (R = 0.0283). Hypothetical frameworks which could well be viable were disclosed like [Ca$_3$Al$_4$F$_{21}$]$^{3-}$ or [NaAl$_2$F$_9$]$^{2-}$ (the latter being obtained as well with Al atoms only but with R = 0.0287, U-9 in Table 2), see the figures below.

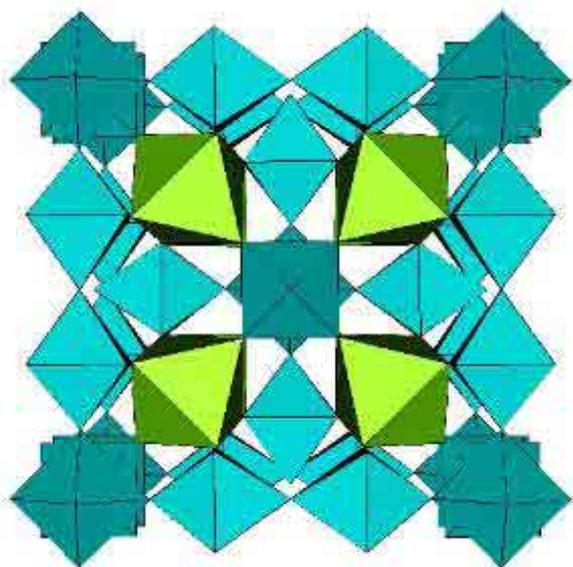

PCOD1000015 - [Ca$_4$Al$_7$F$_{33}$]$^{4-}$.

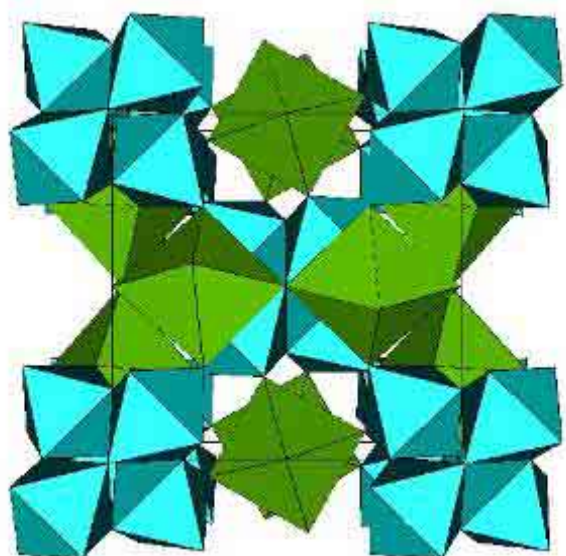

PCOD1010005 - [Ca$_3$Al$_4$F$_{21}$]$^{3-}$.

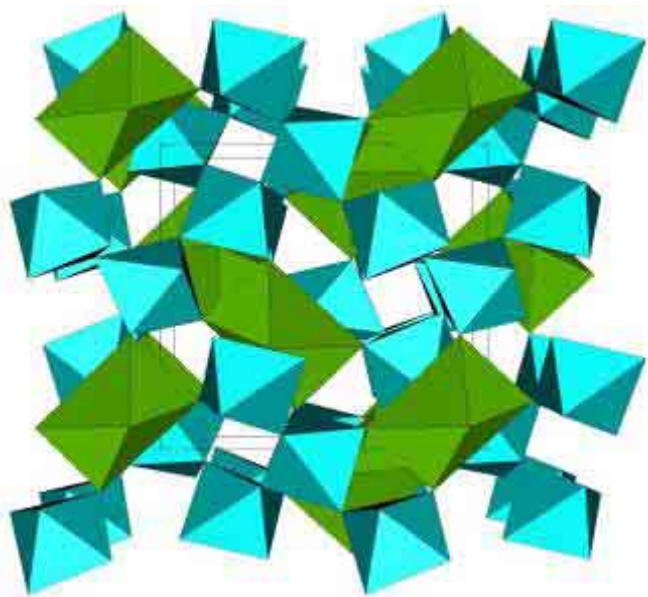

[NaAl$_2$F$_9$]$^{2-}$, Space group P4$_1$32, a = 9.05Å.

**Limitations of the GRINSP software**

There is no way for the random walker in the GRINSP software for exploring something else than the predefined 3, 4, 5 or 6-connected nets leading to corner-sharing polyhedra, in binary ($M_2X_3$, $MX_2$, $MX_3$) or ternary ($M_aM_b'X_c$) compounds. However, it can be easily imagined to introduce more complexity in the predictions, allowing for example to make appear corner-, edge-, and face-sharing polyhedra, altogether, and to propose an automatic way to obtain an electrical neutrality by the detection of holes and the filling of these holes by large cations. It appears clear that the use of bond valence rules would be more efficient when optimizing the final models than the use of simple ideal interatomic distance considerations. There is a project in this direction for improving GRINSP.

**Prediction confirmation**

More difficult even is the prediction of the synthesis conditions for making to appear the predicted crystal structures. However, at least if the chemical composition is more complex than $SiO_2$ or $AlF_3$, one may try the battery of classical methods. If an interesting model is predicted having the $[Ca_3Al_4F_{21}]^{3-}$ formulation, may be it could be really synthesized as $Na_3Ca_3Al_4F_{21}$ or $Li_3Ca_3Al_4F_{21}$, or may be not. We can already be sure that most predictions will be vain, never confirmed, because the synthesis route may depend on a precursor (organometallic, hydrate, amorphous compound) which itself is yet unknown, or because the prediction is simply false. One of the latest discovered $\tau$-$AlF_3$ variety [18] was obtained from the thermolysis of either $[(CH_3)_4N]AlF_4 \bullet H_2O$ or amorphous $AlF_3 \bullet xH_2O$ (x < 0.5). It is a unique example, no other $MX_3$ was found yet to adopt this structure, though there would be no geometrical or lattice energy objection to the existence of some analogous $\tau$-$FeF_3$, $\tau$-$VF_3$ or $\tau$-$CrF_3$ polymorphs. The $[Ca_4Al_7F_{33}]^{4-}$ network proposed by GRINSP in the above table really exists with the $Na_4Ca_4Al_7F_{33}$ formulation. The database of hypothetical zeolites [5] contains more than 100000 proposals, although the number of zeolite different types is less than 200, increasing quite slowly! More modestly, the PCOD [6] will propose probably one thousand selected hypothetical zeolites. For the confirmation of the predictions, we will have to wait for decades or centuries, who knows. Anyway, structure prediction is an unavoidable part of our future in crystallography and chemistry. One can predict also that the accuracy of the structure prediction methods will improve.

**References**

[1] F. C. Hawthorne, *Acta Cryst*. **B50** (1994) 481-510.
[2] C.R.A Catlow & G.D. Price, *Nature* **347**(1990) 243-248.
[3] Computer Modelling in Inorganic Crystallography, C.R.A Catlow (ed), Academic Press, 1997.
[4] W.D.S. Motherwell et al., *Acta Cryst*. **B58** (2002) 647-661.
[5] M.D. Foster & M.M.J. Treacy - Hypothetical Zeolites - http://www.hypotheticalzeolites.net/
[6] A. Le Bail - Predicted Crystallography Open Database - http://www.crystallography.net/pcod/
[7] M.C Payne et al., *Rev. Mod. Phys*. **64** (1992) 1045.
[8] J. D. Gale, *J. Chem. Soc., Faraday Trans.,* **93** (1997) 629-637. http://gulp.curtin.edu.au/

[9] S.M. Woodley, in: Application of Evolutionary Computation in Chemistry, R. L. Johnston (ed), Structure and bonding series, Springer-Verlag **110** (2004) 95-132.
[10] J.C. Schön & M. Jansen, *Z. Krist.* **216** (2001) 307-325; 361-383.
[11] M.W. Lufaso & P.M. Woodward, *Acta Cryst*. **B57** (2001) 725-738.
[12] C. Mellot-Drazniek, J.M. Newsam, A.M. Gorman, C.M. Freeman & G. Férey, *Angew. Chem. Int. Ed*. **39** (2000) 2270-2275; C. Mellot-Drazniek, S. Girard, G. Férey, C. Schön, Z. Cancarevic, M. Jansen, *Chem. Eur. J.* **8** (2002) 4103-4113.
[13] Cerius2, Version 4.2, Molecular Simulations Inc., Cambridge, UK, 2000.
[14] A. Le Bail, *IUCr Comp. Comm. Newsletter* **4** (2004) 37-45. http://www.cristal.org/grinsp/
[15] Ch. Baerlocher, A. Hepp. and W.M Meier, "DLS-76, a program for the simulation of crystal structures by geometric refinement". (1978). Lab. f. Kristallographie, ETH, Zürich.
[16] Database of Zeolite Structures - http://www.iza-structure.org/databases/
[17] G.O. Brunner and F. Laves, *Wiss. Zeitschr*. Techn. Univers. Dresden **20** (1971) 387; W.M,. Meier and H.J. Moeck, *J. Solid State Chem.* **27** (1979) 349-355.
[18] A. Le Bail, J.L. Fourquet & U. Bentrup, *J. Solid State Chem.* **100** (1992) 151-159.